\documentclass[
superscriptaddress,
reprint,
 amsmath,amssymb,
 aps, physrev,
]{revtex4-2}

\usepackage[none]{hyphenat}
\usepackage{graphicx}
\usepackage{microtype} 
\usepackage{hyperref} 
\usepackage{wasysym} 
\usepackage{multirow} 
\usepackage{array} 
\usepackage{makecell} 

\begin{document}

\title{High-Repetition-Rate Laser-Driven Fast Neutron Generation from Planar Liquid Heavy Water Jets}

\author{F. Treffert}
\thanks{These authors contributed equally to this work.}
\email[]{franziska.treffert@focused-energy.co}
\altaffiliation[Now at ]{Focused Energy Inc., San Francisco, CA 94105}
\affiliation{SLAC National Accelerator Laboratory, Menlo Park, CA 94025, USA}
\affiliation{Lawrence Livermore National Laboratory, Livermore, CA 94550, USA}

\author{G. D. Glenn}
\thanks{These authors contributed equally to this work.}
\email[]{gdglenn@sandia.gov}
\affiliation{SLAC National Accelerator Laboratory, Menlo Park, CA 94025, USA}
\affiliation{Applied Physics Department, Stanford University, Stanford, CA 94305, USA}
\affiliation{Sandia National Laboratories, Albuquerque, NM 87123, USA}

\author{C. B. Curry}
\altaffiliation[Now at ]{Pacific Fusion, Fremont, CA 94538}
\affiliation{SLAC National Accelerator Laboratory, Menlo Park, CA 94025, USA}
\affiliation{Department of Electrical and Computer Engineering, University of Alberta, Edmonton, AB, Canada}

\author{D. P. DePonte}
\affiliation{SLAC National Accelerator Laboratory, Menlo Park, CA 94025, USA}

\author{R. C. Hollinger}
\affiliation{Electrical and Computer Engineering Department, Colorado State University, Fort Collins, CO 80521, USA}

\author{G. Jain}
\affiliation{SLAC National Accelerator Laboratory, Menlo Park, CA 94025, USA}
\affiliation{Mechanical Engineering Department, Stanford University, Stanford, CA 94305, USA}

\author{S. Popa}
\affiliation{Extreme Light Infrastructure - Nuclear Physics, ``Horia Hulubei'' National Institute for R\&D in Physics and Nuclear Engineering, Magurele, Romania}
\affiliation{Doctoral School of Physics, University of Bucharest, Magurele, Ilfov, Romania}
\affiliation{Engineering and Applications of Lasers and Accelerators Doctoral School, National University of Science and Technology Politehnica of Bucharest, Bucharest, Romania}

\author{M. Ricks}
\affiliation{SLAC National Accelerator Laboratory, Menlo Park, CA 94025, USA}
\affiliation{Biomedical Physics Program, Stanford University, Stanford, CA 94305, USA}

\author{J. J. Rocca}
\affiliation{Electrical and Computer Engineering Department, Colorado State University, Fort Collins, CO 80521, USA}
\affiliation{Physics Department, Colorado State University, Fort Collins, CO 80523, USA}

\author{B. Sullivan}
\affiliation{Electrical and Computer Engineering Department, Colorado State University, Fort Collins, CO 80521, USA}

\author{D. Ursescu}
\affiliation{Extreme Light Infrastructure - Nuclear Physics, ``Horia Hulubei'' National Institute for R\&D in Physics and Nuclear Engineering, Magurele, Romania}
\affiliation{Doctoral School of Physics, University of Bucharest, Magurele, Ilfov, Romania}
\affiliation{Engineering and Applications of Lasers and Accelerators Doctoral School, National University of Science and Technology Politehnica of Bucharest, Bucharest, Romania}

\author{S. Wang}
\affiliation{Electrical and Computer Engineering Department, Colorado State University, Fort Collins, CO 80521, USA}

\author{G. J. Williams}
\affiliation{Lawrence Livermore National Laboratory, Livermore, CA 94550, USA}

\author{S. Zahedpour}
\affiliation{Electrical and Computer Engineering Department, Colorado State University, Fort Collins, CO 80521, USA}

\author{S. H. Glenzer}
\affiliation{SLAC National Accelerator Laboratory, Menlo Park, CA 94025, USA}
\affiliation{Mechanical Engineering Department, Stanford University, Stanford, CA 94305, USA}

\author{M. Gauthier}
\email[]{gauthier@slac.stanford.edu}
\affiliation{SLAC National Accelerator Laboratory, Menlo Park, CA 94025, USA}

\begin{abstract}
We demonstrate a high-repetition-rate, directional pitcher-catcher neutron source employing a continuously flowing, ambient-temperature liquid jet pitcher and a beryllium catcher target. Proton or deuteron beams are accelerated at 0.5 Hz from water (H\textsubscript{2}O) or heavy water (D\textsubscript{2}O) sheet jets using the $2\omega$ ALEPH laser at Colorado State University. Deuteron-induced neutron generation shows an average flux enhancement in the target normal direction of 13x compared to proton-induced neutron generation, achieving peak fluxes of $7.9\times10^{5}$ neutrons/sr. The experimentally predicted ion-to-neutron conversion efficiency of 1.5$\times 10^{-5}$ for protons and 1.1$\times 10^{-3}$ for deuterons indicates a significant yield contribution from the deuteron breakup reaction. High shot-to-shot stability neutron spectra up to 8.7~MeV ($^9$Be(d,n)$^{10}$B) are observed for more than 200 consecutive shots.
\end{abstract}


\maketitle

\section{Introduction}
Short, intense bursts of fast neutrons are of interest for a wide range of applications, including nondestructive imaging \cite{Mor2009,Li2022}, radiation effects studies \cite{LoPresti2020}, medicine \cite{Gray1943,Jones2020}, and fusion energy \cite{Perkins2000,Higginson2010}. These applications are best served by sources delivering multi-MeV neutron energies, high peak fluxes, and high average fluxes enabled by high repetition rate (HRR, $\gtrsim$1~Hz) operation. Nanosecond-scale neutron pulse durations are also beneficial for certain applications. Conventional sources of fast neutrons such as nuclear reactors or spallation sources typically fulfill only some of these requirements while also presenting significant challenges for many of the most promising applications due to their spatial footprint, cost, and limited flexibility in tailoring neutron beam parameters. Recent advances in high power, HRR laser technology \cite{Sistrunk2019, Tamer2024}, combined with the development of continuously refreshing targets \cite{Curry2020a, Condamine2021,Treffert2022,Treffert2022a,Faubel2025,Miao2025,Drobniak2026}, now enable laser-driven neutron sources to deliver MeV neutron beams with very high peak and average fluxes and nanosecond pulse durations at Hz-scale shot rates.

Laser-driven neutron sources based on short-pulse, high-intensity laser interactions predominantly generate neutrons through three different mechanisms: deuterium-deuterium (d(d,n)$^3$He, or D-D) fusion, proton- (p,n) or deuteron- (d,n) induced nuclear reactions, or photonuclear ($\gamma$,xn) reactions. D-D fusion sources are typically based on direct irradiation of a target with a high-intensity laser; fusion events occur both in the bulk of the target and in the  subsequent explosion of the hot target plasma \cite{Norreys1998,Ditmire1999,Disdier1999,Curtis2021}. A small number of studies have also investigated a two-target approach to D-D fusion neutron sources in which deuterons are accelerated from a primary target and induce D-D fusion reactions in a secondary target such as a deuterated plastic (CD) foil  \cite{Maksimchuk2013,Kar2016,Wang2025}. D-D fusion neutron sources produce quasi-monoenergetic neutron spectra centered at 2.45~MeV, characteristic of the d(d,n)$^3$He reaction, but their practical applications have historically been limited by their low yields, isotropic emission, and limited tunability. Some studies have demonstrated D-D neutron generation at repetition rates up to kHz using liquid targetry \cite{Karsch2003,Hah2016,Knight2024}, but the average neutron fluxes achieved have been limited by the modest laser energies available at such high repetition rates.

The ``pitcher-catcher'' scheme of laser-driven neutron generation has attracted significant interest due to its potential to generate broadband fast neutron spectra with high flux \cite{Lancaster2004,Yang2004,Higginson2011,Willingale2011,Roth2013,Treffert2021,Yogo2023,Millan-Callado2026,Pomerantz2014,Qi2019,Li2022b,Vallieres2025}. In this approach, a high-intensity laser irradiates a ``pitcher'' target and accelerates a particle beam composed of ions or electrons to MeV energies. The laser-driven particle beam is then incident on a ``catcher '' target, the composition of which is dependent on the incident particle type. 

Neutron generation using electron beams accelerated either via direct laser acceleration \cite{Pomerantz2014,Qi2019} or laser wakefield acceleration (LWFA) \cite{Li2022b,Vallieres2025} has recently shown promising results. In these interactions, neutrons are generated in a high-Z (e.g. W) catcher via photonuclear fission reactions driven by bremsstrahlung emission from the energetic electron beam. This approach to laser-driven neutron generation relies on the giant dipole resonance at high photon energies (few tens of MeV) causing the neutron spectrum to be less tunable than for sources driven by ions. Additionally, the need for high energy photons and the low conversion efficiency from photons to neutrons (${\sim}$10$^{-4}$, approximately comparable to the proton-to-neutron conversion efficiency in ion-driven pitcher-catcher interactions) leads to a significantly higher flux of gamma rays. This intense gamma ray background can saturate neutron detectors and reduce signal-to-noise levels. In sample irradiation experiments, these gammas may also trigger undesirable reactions in the material sample. Additionally, the intense flux of ionizing radiation may interfere with the operation of electronic equipment in the vicinity of the laser target chamber.

In the ion beam-driven approach, a high-intensity laser irradiates a pitcher target and accelerates protons or deuterons to MeV energies via mechanisms such as target normal sheath field acceleration (TNSA) \cite{Snavely2000,Wilks2001}, breakout afterburner \cite{Yin2006}, or similar. The laser-driven ion beam is incident on a catcher typically composed of Li or Be. Nuclear reactions between the protons or deuterons and the target nuclei, most notably the ``deuteron breakup'' reaction, produce a quasi-directional, high-flux, broadband neutron beam. In contrast to the ``complete fusion'' reaction, which captures the incident nucleon and subsequently releases a neutron, during the deuteron breakup reaction the deuteron disintegrates into a proton and a neutron in the atomic potential of converter atoms and generates an additional forward-directed neutron beam due to conservation of momentum. The cross-sections of the complete fusion and deuteron breakup reactions are maximized between several MeV and several tens of MeV, facilitating the production of neutron peak energies of many tens of MeV \cite{Roth2013}. Importantly, the neutron spectra obtained from such sources are significantly shaped by the incident ion spectrum. This can provide control over the spectral shape of the neutron beam, particularly when accessing ion acceleration regimes that provide narrowband ion energy spectra. The highest neutron fluxes from pitcher-catcher sources have been obtained in single-shot, high power laser experiments \cite{Kleinschmidt2018, Yogo2023} with correspondingly low average flux. Recent work has integrated HRR lasers and primary targets to operate ion-driven pitcher-catcher neutron sources at Hz levels or beyond, achieving significantly higher average fluxes at moderate peak fluxes \cite{Lelievre2024,Osvay2024,Stuhl2025}.

In this work, we describe a 0.5~Hz neutron source driven by proton and deuteron beams accelerated from liquid water (H\textsubscript{2}O) and heavy water (D\textsubscript{2}O) sheet jet targets \cite{Treffert2022, Treffert2022a} using the $2\omega$ ALEPH laser at Colorado State University \cite{Wang2017}. We implemented a beryllium converter with an on-axis pinhole, allowing direct correlation between ion and neutron spectra on each shot and the calculation of an experimental conversion efficiency from ions to neutrons. Peak fluxes for neutron generation using a D\textsubscript{2}O target reached up to 5~$\times$~10\textsuperscript{5} neutrons/MeV/sr on selected shots, presenting an increase of more than an order of magnitude compared to neutron fluxes achieved using proton beams accelerated from an H\textsubscript{2}O target. Our experimental observations are compared to Monte Carlo particle transport code simulations using the measured ion beam parameters and we find good agreement between the experimental and modeled neutron spectra.

\begin{figure}
    \centering
    \includegraphics[width=0.7\linewidth]{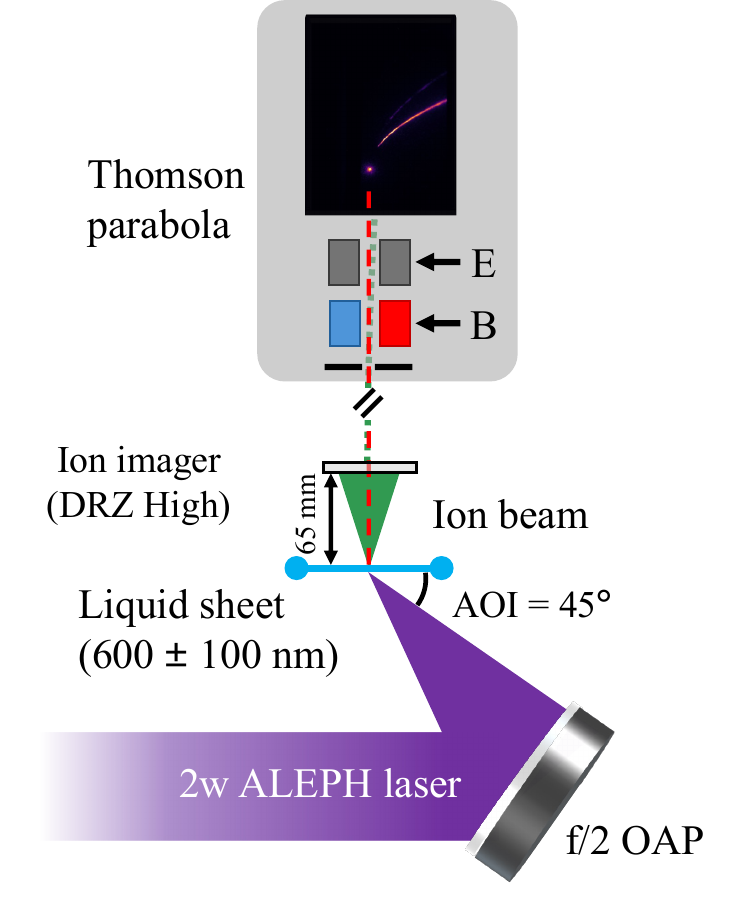}
    \caption{A schematic illustration of the ion characterization phase of the experiment. Protons or deuterons were accelerated from the interaction of the high-intensity $2\omega$ ALEPH laser (400~nm, 4.8~J, 45~fs, 0.5~Hz, up to $8.5\times10^{20}$~W/cm\textsuperscript{2}) with a 600~nm thick H\textsubscript{2}O or D\textsubscript{2}O liquid sheet jet. The ion beam was characterized spatially and spectrally using a scintillator-based ion imager and a Thomson parabola (TP) aligned with the nominal target normal axis.}
    \label{fig:ion_setup}
\end{figure}

\section{Experiment Setup}
The $2\omega$ ALEPH laser (400~nm, 45~fs, 4.8~$\pm$~0.3~J, $8.5\times10^{20}$~W/cm\textsuperscript{2}) was focused onto the target at a 45$^\circ$ angle of incidence using an f/2 off-axis parabola (OAP). The laser was s-polarized with respect to the target plane. The target consisted of planar liquid H\textsubscript{2}O or D\textsubscript{2}O sheets produced using tungsten microfluidic nozzles \cite{Treffert2022a}. During the experiment, the liquid jet operated at typical pump pressures of 500--600~psi and flow rates of approximately 5.5~mL/minute. Characterization of the thickness of the liquid sheet using white light interferometry set up in reflection prior to selected shots indicated a liquid jet thickness at the laser-target interaction location of 600~$\pm$~100~nm. The target thickness can also be estimated using an established scaling relation for the thickness of planar liquid sheets produced using converging nozzles with the geometry used here,
\begin{equation}
    h = 0.84\frac{d w_0}{x}\,,
    \label{eq:sheet_thickness}
\end{equation}
where $h$ is the thickness of the liquid sheet, $x$ is the distance from the nozzle output in the plane of the sheet, and $d$ and $w_{0}$ are the depth and width of the nozzle channel respectively \cite{Ha2018,Crissman2022}. In the experiment described here the laser-target interaction was located at a distance of 3.5~mm from the output surface of the metal nozzle, which had a channel width of 100~\textmu m and a depth of 25~\textmu m. With these parameters Eq.~\ref{eq:sheet_thickness} predicts a sheet thickness of 636~nm, consistent with the in-situ target thickness measurement. 

Several measures were employed to mitigate the impact of the liquid jet target on the vacuum environment in the target chamber and the laser compressor. The liquid jet was intercepted by a heated catcher approximately 5~mm below the interaction region, which evacuated the liquid to a reservoir outside the chamber using a combination of vapor pressure and gravity. We also implemented a LN\textsubscript{2} cold finger inside the target chamber that was replenished at regular intervals during liquid jet operation. These efforts ensured that the pressure inside the interaction chamber during jet operation was typically ${\sim}1\times10^{-3}$~mbar. As described in Ref.~\citenum{Treffert2022a}, the interaction chamber was isolated from the laser compressor with a 0.5~mm thick uncoated glass pellicle to protect the compressor gratings from vapor generated by the liquid jet. Reflection losses at the pellicle surfaces have been considered for all on-target laser energy values given in this work.

\section{High repetition rate ion beam generation}
\label{sec:ions}
\subsection{Ion Diagnostics}
Ions accelerated by the high intensity laser interaction with the liquid sheet were characterized both spectrally and spatially. The absolute energy spectra of the laser-driven ion beams were measured using a differentially-pumped Thomson parabola spectrometer (TP) with a line of sight along the nominal target normal axis (depicted in \autoref{fig:ion_setup}) and an angular acceptance of $6.9\times10^{-8}$~sr. Further details on the construction of the TP are provided in Ref.~\citenum{Nedbailo2023}. Ions were detected using a microchannel plate (MCP) detector coupled to a phosphor screen to visualize the ion signal generated in the MCP with high efficiency. Light emitted by the phosphor screen was recorded using a thermoelectrically cooled sCMOS camera (Thorlabs Quantalux CC215MU). The TP was absolutely calibrated using a piece of slotted solid-state CR-39 nuclear track detector \cite{Harres2008} which was attached to the front of the MCP detector for a single shot. Detection of the ion flux simultaneously on the CR39 and the MCP assembly provided an absolute calibration via direct comparison between the camera counts on the image of the phosphor screen and the absolute ion numbers obtained by etching and counting the ion tracks on the CR-39.

The ion beam spatial profile was measured using a single layer of DRZ High GOS scintillator (Mitsubishi Chemical) \cite{Schilz2024} placed at a distance of 65~mm from the interaction, referred to hereafter as the ``ion imager.'' A slit from one vertical edge to the center of the ion imager along its center axis allowed the unperturbed ion beam spectrum to be measured using the TP. The scintillator active surface, facing the target, was covered in a 25~\textmu m-thick layer of aluminum to eliminate potential contributions from heavy ions and reduce background levels. Modeling of the filter and scintillator structure using SRIM \cite{Ziegler2010} indicated that the minimum ion energy detected by the scintillator was approximately 1.7~MeV for protons and 2.2~MeV for deuterons. For the energy spectra generated here, all ions above these energies were stopped in the scintillator. Light produced by the scintillator was collected using a CCD camera (Basler acA1600-20gm) equipped with bandpass filtering for the central emission wavelength (545~nm) of the scintillator.

\subsection{Ion beam properties}
Spectral and spatial characterization of the laser-driven ion beams was essential to quantitatively predict neutron fluxes and investigate the dependence of the neutron beam parameters on the ion beam properties. Ion spectra from the TP were used to determine the maximum ion energy E\textsubscript{Max} and the peak ion flux $\Phi_\text{Max}$ along the nominal target normal direction. The ion imager was used to measure the typical pointing direction of the ion beam and the full width at half maximum (FWHM) divergence for the portion of the ion beam with energies high enough to penetrate into the scintillator (E\textsubscript{Min}). The beam pointing and divergence were extracted from the raw data via Gaussian fits to vertically- and horizontally-averaged ion beam spatial profiles. The pointing values were calculated with respect to the nominal target normal axis in the plane of the laser. Here, we present ion beam data collected immediately prior to the implementation of the neutron converter for both protons and deuterons. The ion properties determined using the TP and ion imager data can therefore be considered to be representative of the ion beam properties in the subsequent neutron generation experiment. We summarize the key properties of the proton and deuteron beams in \autoref{tab:ion_comparison}. The proton measurements are averaged over 116 shots for the TP and 94 shots for the ion imager due to trigger failures in the ion imager camera, and the deuteron measurements are averaged over 61 shots for the TP and 44 shots for the ion imager. All data were obtained continuously at a repetition rate of 0.5~Hz.

\begin{table}
\caption{Key features of the proton (H\textsubscript{2}O target) and deuteron (D\textsubscript{2}O target) beams characterized immediately before neutron generation experiments.}
\label{tab:ion_comparison}
\begin{ruledtabular}
\begin{tabular}{cccc}
Diagnostic & Parameter & Protons & Deuterons \\
\hline
\multirow{2}{*}{\makecell{Thomson\\parabola}}
  & $E_{\mathrm{max}}$ (MeV) & $3.5 \pm 0.3$ & $4.1 \pm 0.4$ \\
  & \makecell{$\Phi_{\mathrm{max}}$\\($10^{11}$/MeV/sr)} & $12.6 \pm 3.5$ & $3.8 \pm 1.2$ \\
\hline
\multirow{3}{*}{\makecell{Ion\\imager}}
  & $E_{\mathrm{min}}$ (MeV) & 1.7 & 2.2 \\
  & Pointing ($^\circ$) & $-3.3 \pm 1.6$ & $-12.1 \pm 1.4$ \\
  & \makecell{FWHM\\Divergence ($^\circ$)} & $12.4 \pm 3.5$ & $11.7 \pm 1.3$ \\
\end{tabular}
\end{ruledtabular}
\end{table}

We observe that the peak proton flux was a factor of 3.3 higher than the peak deuteron flux and showed similar relative flux stability. While the deuteron beam achieved a higher total energy (E\textsubscript{Max}), its energy per nucleon was lower than that of the proton beam. The pointing stability for both ion beams was comparably high, with shot-to-shot variations in the ion beam pointing of only 1--2$^\circ$. However, the deuteron beam was directed significantly further from the nominal target normal direction than the proton beam. This may indicate that the liquid sheet had rotated prior to the acquisition of this dataset, since for the laser and target parameters used here we anticipate that the ion beams were accelerated normal to the target rear surface via sheath acceleration. In both cases the ion beam divergence was relatively low for ion beams accelerated from sheath acceleration and was stable at slightly more than 10$^\circ$ \cite{Roth2016}.

\begin{figure}
    \centering
    \includegraphics[width=\linewidth]{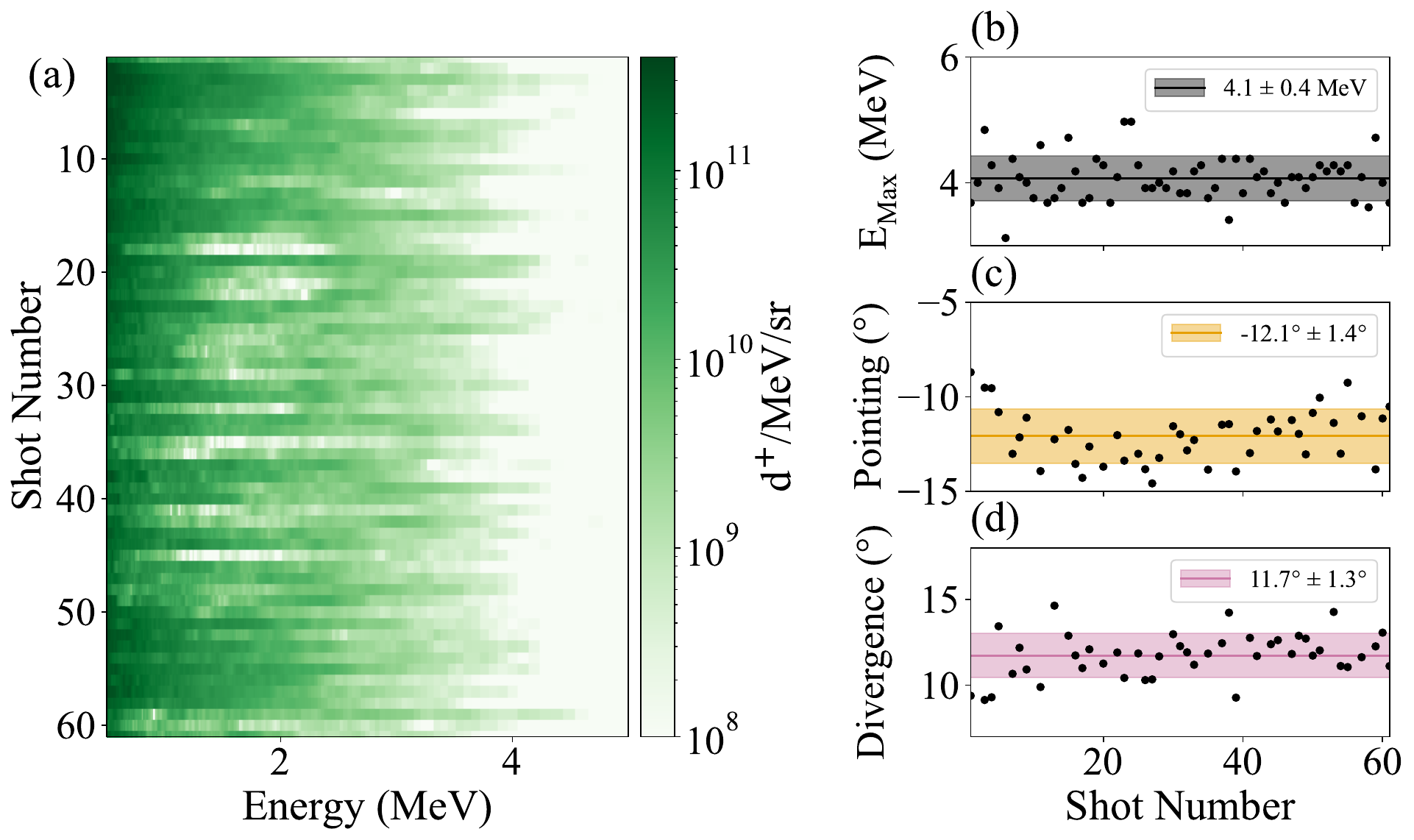}
    \caption{(a) Absolute deuteron energy spectra extracted from the TP across 61 shots obtained at 0.5~Hz. (b) The maximum deuteron energy (E\textsubscript{Max}) for each shot is shown in (a), with an average E\textsubscript{Max} of 4.1~MeV. (c) The pointing direction with respect to the nominal target normal direction shows an average offset of -12.1$^\circ$. (d) The average full width at half maximum (FWHM) divergence of the deuteron beam is 11.7$^\circ$. The data shown in (c) and (d) were acquired from 44 shots on which the ion imager camera successfully triggered.}
    \label{fig:csu_d_beam}
\end{figure}

The shot-to-shot performance of the deuteron beam with respect to these performance metrics is shown in more detail in \autoref{fig:csu_d_beam}. The waterfall plot in \autoref{fig:csu_d_beam}(a) shows that the deuteron spectra were generally quite stable shot-to-shot. While individual shots show local variations in the ion spectral shape, particularly at ion energies from 1--2~MeV, the peak fluxes are generally above $10^{10}$~d\textsuperscript{+}/MeV/sr for ion energies below 2~MeV and show a smooth falloff with increasing energy. The results improve on the deuteron beam stability and flux presented in our previous work \cite{Treffert2022} while attaining similar maximum ion energies across a much higher fraction of the shots. The deuteron maximum energy, plotted over the shot series in \autoref{fig:csu_d_beam}(b), was found to be $4.1\pm0.4$~MeV, implying that the shot-to-shot stability of the maximum energy was approximately 10\% (relative standard deviation). The deuteron beam also showed high spatial stability, with typical variations in the pointing and divergence of only approximately 1$^\circ$. We do not observe overall drifts in any of these parameters throughout the shot series, suggesting that the liquid sheet jet thickness, position, and orientation remained stable and the tungsten nozzle was not degraded by close proximity to 0.5~Hz high-intensity laser-matter interactions.

\section{High repetition rate neutron beam generation}
\label{sec:neutrons}
After characterizing the ion source, we modified our experimental configuration to generate neutrons in the pitcher-catcher geometry. The ion imager was replaced with a \textsuperscript{9}Be converter (1~mm thick, \diameter25.4~mm) mounted in a \diameter25.4~mm lens tube at a distance of 39~mm from the liquid jet target. A 10~\textmu m-thick layer of aluminum was placed in front of the converter to protect it from ablation by scattered laser light and heavy ions. The converter featured an on-axis \diameter1.6~mm through-hole that maintained the TP line of sight for continued on-shot measurements of the ion energy spectrum. The converter and ion imager were mounted on the same motor stage, allowing us to replace the ion imager with the converter in less than a minute without breaking vacuum or interrupting the operation of the liquid jet target. 

\subsection{Neutron diagnostics}
\begin{figure}
    \centering
    \includegraphics[width=\linewidth]{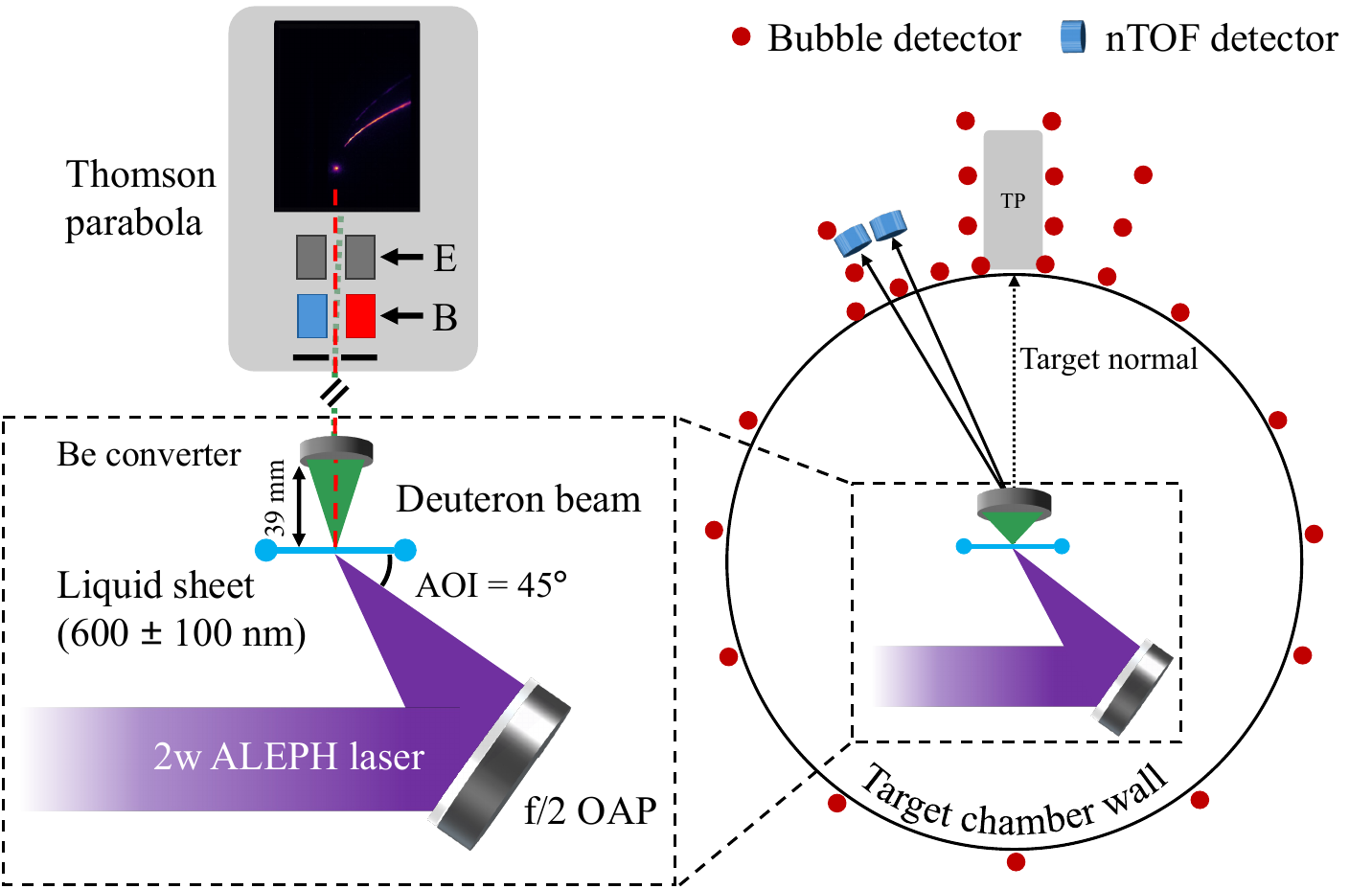}
    \caption{A schematic illustration of the neutron generation phase of the experiment. Deuterons were accelerated from the high-intensity laser interaction and impinged on a Be converter placed 39~mm from the interaction. The converter featured an on-axis 1.6~mm diameter aperture that allowed the TP to continue to monitor the ion beam energy spectrum during neutron generation. The neutrons' spatial distribution and energy spectrum were characterized using both neutron bubble detectors (red dots) and neutron time-of-flight detectors (blue cylinders) distributed around the target chamber as illustrated on the right.}
    \label{fig:neutron_setup}
\end{figure}

Neutron detectors located outside the target chamber characterized the spatial distribution and energy spectrum of the neutrons as shown in Figure \ref{fig:neutron_setup}. The spatial distribution was measured with 26 calibrated high sensitivity (20-40 bubbles/mrem) bubble detector personal neutron dosimeters \cite{Ing1997} (BD-PND, Bubble Technology Industries) placed at a variety of distances and angles around the target chamber with higher coverage in the target normal direction. These detectors are sensitive to neutron energies from roughly 100~keV to 10~MeV and rely on the detection of neutrons through energy deposition within superheated liquid dispersed in a polymer gel and subsequent bubble formation \cite{Lewis2012,Jung2013}. For the conversion of the bubble detector measurements to absolute neutron numbers, we used the response curve in Ref.~\citenum{Jung2013}. The uncertainty of each individual bubble detector measurement is taken to be 20\% following previous response characterization results \cite{Rosenstock1995,Ing1997}.

Two neutron time-of-flight (nTOF) detectors based on plastic scintillators (Eljen EJ-228 or EJ-232Q) coupled to photomultiplier tubes (Hamamatsu H2431-50) measured the energy spectrum of the forward-directed neutron beam. The nTOFs were located at angles of -22 and -27 degrees from the nominal target normal direction at a distance of 1.1~m from the converter and were shielded with 5 cm of Pb along the line of sight. The nTOF detector located at -27$^\circ$ was absolutely calibrated at a photomultiplier operating voltage of -2~kV using a deuteron-deuteron (D-D) fusion neutron source of known flux. During the experiment this detector was operated at -1.8~kV. The corresponding calibration value was calculated using the amplification ratio between the calibration voltage and the operating voltage according to the gain curve of the H2431-50 photomultiplier. The second nTOF detector (-22$^\circ$) was cross-calibrated with the absolutely calibrated nTOF using data from 204 consecutive shots measuring deuteron-driven neutrons produced in the Be converter, as described in Section~\ref{sec:neutrons}. To obtain the cross-calibration, we calculated a calibration value for each shot, then took the average across shots as the overall calibration value. The uncertainty in the calibration value was the standard deviation across shots. This approach assumes that the neutron flux does not vary significantly over the 5$^\circ$ separation between the detectors such that both detectors measure comparable neutron yields within the experimental uncertainty. The neutron yields per steradian derived from the two nTOFs were then compared with the measurement from the closest bubble detectors, showing good agreement within experimental uncertainties.

In the following we examine the spatial distribution of the laser-generated neutrons using bubble detector data and shot-integrated nTOF data. Then, we describe the energy spectrum of the neutrons and their shot-to-shot stability as characterized by the nTOF detectors. Finally, we compare neutron generation with and without the converter for incident protons and deuterons.

\subsection{Neutron angular distribution and source determination}
\begin{figure*}[htb]
    \centering
    \includegraphics[width=\linewidth]{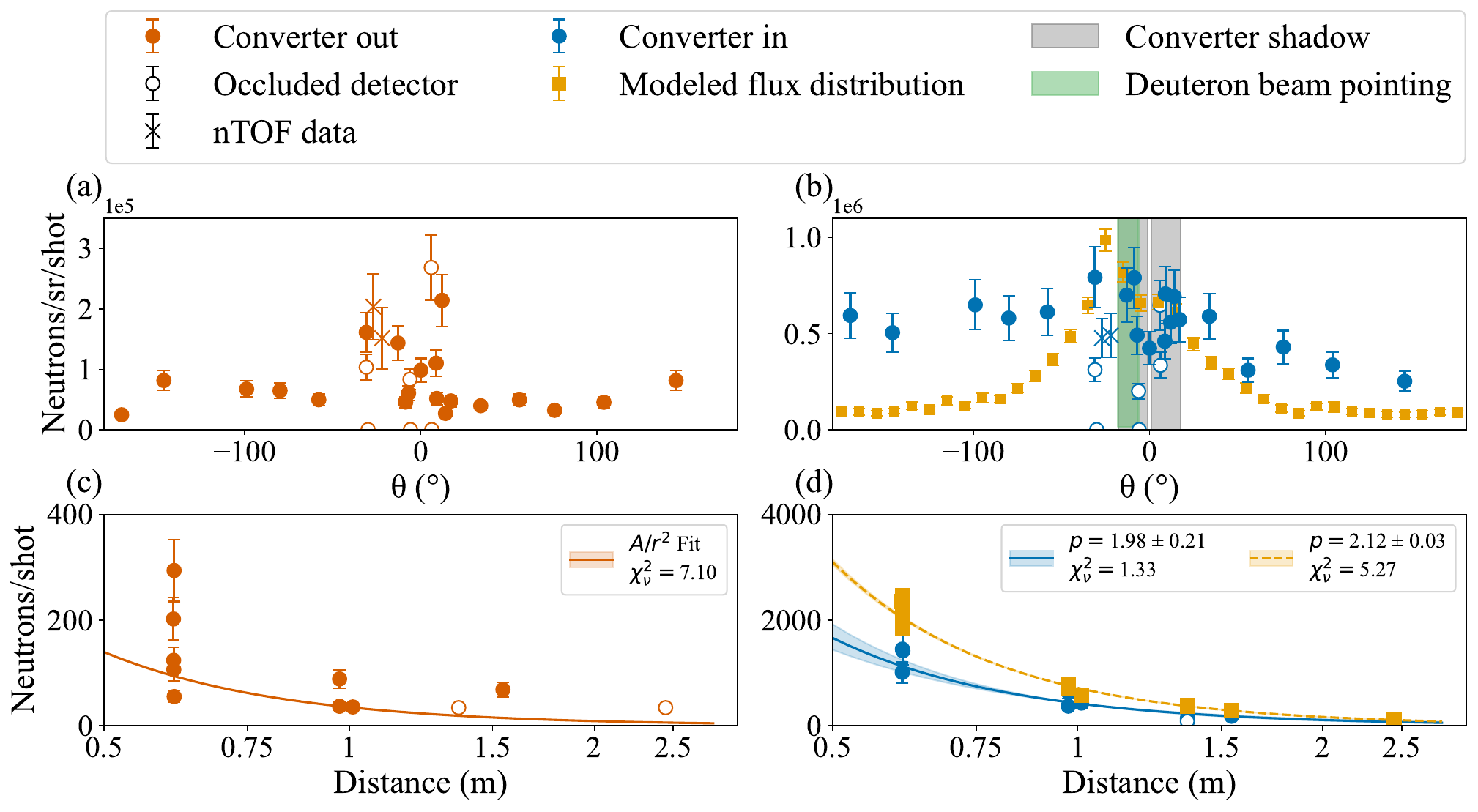}
    \caption{(a) Average neutron flux measured by bubble detectors placed at various angles for an incident deuteron beam without the converter in place. (b) Average neutron flux measured by bubble detectors placed at various angles for an incident deuteron beam with the converter in place. The angular shadow of the converter and the locations of the nTOF detectors are identified. Due to the nTOF calibration method, both nTOFs measure close to the same flux for the converter-in case. The results of a PHITS simulation (yellow squares) show a much narrower angular distribution but comparable peak fluxes. (c) The average neutron flux measured without the converter as a function of bubble detector distance from the converter. A fit to $A/r^{2}$ shows $\chi^{2}_{\nu}$ significantly larger than 1, consistent with the prediction that these neutrons were produced from an extended source rather than a point source. (d) The average neutron flux measured with the converter as a function of bubble detector distance from the converter from both the experiment and the PHITS simulation. A power-law fit to these data shows $p\approx2$ in both cases, confirming that a point-source model is a much better fit to the data than in (c). These results confirm the converter was the primary source of the forward-directed neutron beam.}
    \label{fig:neutron_spatial}
\end{figure*}

The spatial distribution of the neutron emission for deuteron beams incident on the ion imager (``converter out'') or the converter (``converter in'') is shown in \autoref{fig:neutron_spatial}(a) and \ref{fig:neutron_spatial}(b) respectively. In these plots, the neutron flux measured on each bubble detector or nTOF is shown as a function of the detector angular position with respect to the nominal target normal direction. An angular position of 0$^\circ$ corresponds to the nominal target normal direction, with positive values corresponding to detector positions clockwise from the nominal target normal as shown on the right-hand side of \autoref{fig:neutron_setup}. We further distinguish between bubble detectors that had an unobstructed line of sight to the converter through the target chamber wall (solid points) and bubble detectors whose line of sight was occluded by a detector or large piece of optomechanical hardware in addition to the target chamber wall (empty data points). We expect the flux measured on the latter detectors to be reduced due to neutron attenuation by the occluding hardware. To compare the converter-out and converter-in cases, we treat the geometric center of the converter as the neutron source for both cases and calculate the neutron flux (neutrons/steradian/shot) for all detectors accordingly.

The average neutron flux inferred by the bubble detector measurements for the converter-out case was $(7.5\pm 6.2)\times10^{4}$~n/sr/shot. In this configuration, the neutron flux was on average constant across all angles, with slightly elevated values in the target normal direction. This is the result of the deuteron beam being stopped not by the converter, but by the vacuum chamber wall in the target normal direction. In this case, since the source of neutrons was predominantly located in the chamber wall directly in front of the bubble detectors mounted close to target normal, we only see elevated numbers at smaller angles. At larger angles we observe a lower, approximately constant flux across all angles. This can be explained by neutron generation at multiple locations in the target chamber, i.e. wherever the ion beam encountered an object, as well the much larger size of these different sources. Consequently, the neutron flux inferred above is likely different from the true neutron flux generated in the converter-out configuration. Given the variety of neutron sources in this configuration, however, a detailed analysis of the sources of neutrons in the converter-out case is outside of the scope of this work. 

When the converter was inserted, the average neutron flux increased by a factor of 6 to $4.8\times10^{5}$~n/sr/shot, with a peak flux measured by a single detector of $7.9\times10^{5}$~n/sr/shot. Spatially integrating these measurements yields an average total neutron yield of $6.0\times10^{6}$~neutrons/shot or $1.25\times10^{6}$~neutrons/shot per joule of incident laser energy. While the highest neutron fluxes for this configuration were measured in the shadow of the converter (gray shaded area in \autoref{fig:neutron_spatial}(b)), substantial neutron fluxes were also observed at larger negative angles. Since the deuteron beam was measured to point in the negative direction prior to the insertion of the converter (see Section~\ref{sec:ions}, indicated by the green shaded region in \autoref{fig:neutron_spatial}), these neutrons were potentially generated from low-energy portions of the deuteron beam which missed the catcher and generated neutrons in the chamber wall or by neutrons from the converter scattered away from the target normal axis by hardware located in the target chamber. Finally, we also observe a higher number of neutrons at larger negative angles than larger positive angles, which we associate to the larger amount of hardware present in the chamber in this direction. This hardware may have scattered neutrons generated in the converter towards bubble detectors placed at negative angles, creating the constant neutron flux observed at larger negative angles in contrast to the trend of decreasing neutron flux toward larger positive angles.

Characterization of the neutron spatial distribution with the converter in place provides an opportunity to verify that neutrons are predominantly generated in the converter, rather than through interactions with other hardware inside the interaction chamber or in the chamber wall. To account for the directional nature of neutron sources generated in the pitcher-catcher geometry, we only include bubble detectors in the shadow of the converter ($\pm 17.7^\circ$ from target normal), estimated using the rear surface cross-sectional area of the converter disk. When neutrons are predominantly generated in the converter, the flux measured by each bubble detector should approximately scale as $1/r^{2}$ with the distance from the converter $r$, as the converter will approximate a point source on the scale of the detector distances. To assess how well the neutron emission resembled that of a point source, the bubble detector fluxes for the converter-in case were fit to a power-law dependence of the form
\begin{equation}
    F = \frac{A}{r^{p}}\,,\label{eq:powerfit}
\end{equation}
where $F$ is the neutron flux and $A$ and $p$ are fit parameters. The data and fits are presented in \autoref{fig:neutron_spatial}(c--d). When the converter was inserted, the best fit yielded $p=1.98\pm0.21$ with$\chi^{2}_\nu=1.33$, indicating that these data were described by a $1/r^{2}$ dependence within experimental uncertainty. By contrast, a fit of the converter-out data (Fig.~\ref{fig:neutron_spatial}(c)) to $F=A/r^{2}$ yielded $\chi^{2}_{\nu}=7.10$, suggesting that those data were not well-described by a $1/r^{2}$ dependence. This observation is consistent with the absence of the converter and our inference that the neutrons measured in this configuration were predominantly generated in the chamber wall or other components in the vacuum chamber and therefore emitted from an extended source. 

We also modeled the neutron spatial distribution using the Monte Carlo particle and heavy ion transport code PHITS (v3.350) \cite{Sato2024}. The source was modeled using the shot-averaged energy spectrum measured by the TP. The simulated deuteron beam was emitted from a point source at the location of the laser-target interaction with an $11.7^\circ$ FWHM divergence and a beam axis tilted $-12.1^\circ$ from target normal, matching the measured divergence and pointing (\autoref{tab:ion_comparison}). $3\times10^{10}$ deuterons were used in the simulated deuteron beam based on a scan over deuteron number to obtain a best match to the experimental neutron data. The simulated deuteron beam was incident on a converter matching the experimental geometry: a 1~mm thick, \diameter25.4~mm $^{9}$Be disk with a 1.6~mm on-axis aperture and a 10~\textmu m thick layer of aluminum on the front surface. The simulation geometry also included a concentric SS304 shell with the dimensions of the target chamber wall, but omitted the remainder of the in-chamber hardware for simplicity. Deuteron reactions on $^{9}$Be used evaluated data up to 20~MeV/nucleon \cite{Iwamoto2023} supplemented by the INCL cascade and evaporation models and including deuteron breakup. Simulated yields were scaled to absolute neutrons per shot using the measured, shot-averaged absolute deuteron spectrum. The same source was scored in three geometries matching the diagnostics: a ring of scoring volumes at 10$^\circ$ increments around the chamber wall for the angular distribution shown in \autoref{fig:neutron_spatial}(b), volumes at the measured bubble detector positions for the $1/r^{2}$ falloff shown in \autoref{fig:neutron_spatial}(d), and volumes matching the locations of the nTOFs for the neutron spectra shown in \autoref{fig:neutron_energy_spectra}(b).

We observe a comparable peak flux for simulations and experimental measurements of neutron generation in the converter, as shown in \autoref{fig:neutron_spatial}(b). The experimentally-detected flux was observed to be higher than the modeled flux at larger negative and positive angles. As discussed previously, this is likely due to the fact that the setup used for the simulations did not take into account any hardware placed in the chamber and therefore likely underestimated the influence of neutron scattering from this hardware. We also fit the simulated bubble-detector data to Eq.~\ref{eq:powerfit} as above and find a best-fit value of $p=2.12$, in fair agreement with the experimental data (Fig.~\ref{fig:neutron_spatial}(d)). The larger $\chi^{2}_{\nu}$ from the simulated dataset compared to the experiment is the result of its smaller statistical uncertainties. This confirms the converter as the primary source of the target normal-directed neutrons both experimentally and numerically.

\subsection{Simultaneous on-shot deuteron and neutron spectra at high repetition rate}
\begin{figure}
    \centering
    \includegraphics[width=\linewidth]{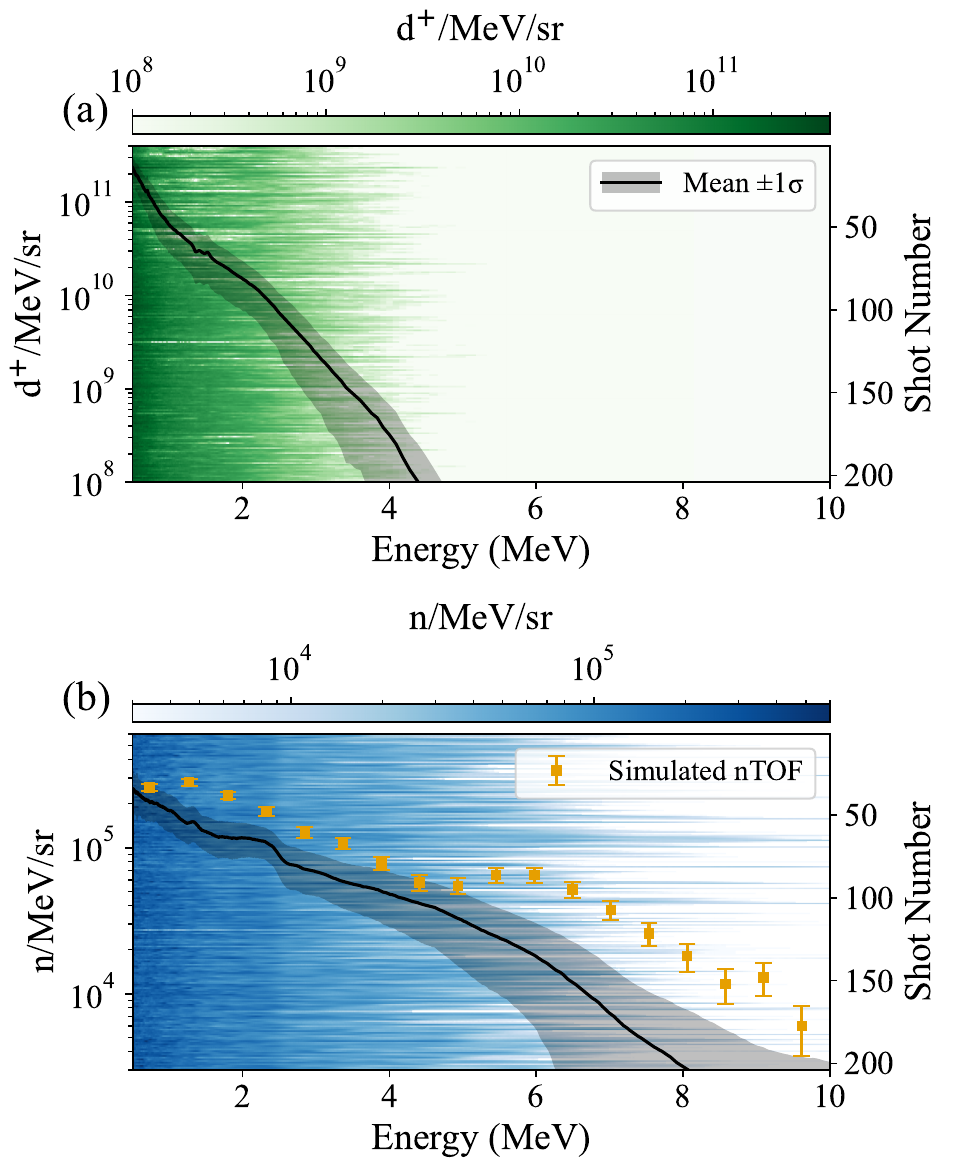}
    \caption{(a) Absolute deuteron energy spectra obtained from the TP over 204 consecutive shots at 0.5~Hz. (b) Absolute neutron energy spectra obtained from the nTOF at -27$^\circ$ over the same set of shots. Both plots also include energy spectra averaged over the shot series, with the shaded area representing one sample standard deviation. Both spectra show a smooth exponential dependence, with a higher cut-off energy for the neutron spectrum due to the positive Q-value of the $^9$Be(d,n)$^{10}$B reaction. The simulated neutron spectrum, shown in yellow, shows similar flux values compared to the measured spectrum for neutron energies below 5~MeV and elevated flux values above that energy.}
    \label{fig:neutron_energy_spectra}
\end{figure}

The on-axis aperture in the converter allowed us to continue to monitor the deuteron energy spectrum on each shot even after the converter was inserted, enabling shot-by-shot comparison to the corresponding neutron energy spectra measured with the nTOF detectors. Deuteron and neutron energy spectra measured in 204 consecutive high-intensity laser shots performed at 0.5~Hz are shown in \autoref{fig:neutron_energy_spectra}. The single-shot deuteron spectra, shown in \autoref{fig:neutron_energy_spectra}(a), exhibit similar energy spectra and stability to those shown in \autoref{fig:csu_d_beam}(a), with an average maximum energy of $4.1\pm0.5$~MeV and a peak flux of $(3.0\pm0.7)\times10^{11}$~d\textsuperscript{+}/MeV/sr. The waterfall plot shows occasional shots where the quality of the interaction decreases and the flux at or below 2~MeV drops significantly. Overall, however, the deuteron maximum energy and flux is quite consistent shot-to-shot. The average deuteron spectrum, which is overlaid on \autoref{fig:neutron_energy_spectra}(a) along with the $\pm1\sigma$ bounds, shows the smooth exponential dependence of flux on particle energy typical of sheath-accelerated ion beams \cite{Roth2016}.

The neutron spectra obtained from the nTOF detector at -27$^\circ$ from the nominal target normal direction are shown in \autoref{fig:neutron_energy_spectra}(b). While these spectra are less stable than the deuteron spectra for neutron energies above 4~MeV, they remain consistent shot-to-shot, especially at the lower neutron energies, which are generally insensitive to the flux variations at lower energies exhibited by the deuteron beam. We observe a strong correlation between total neutron flux and total ion flux and a less pronounced correlation between ion cut-off energy and neutron cut-off energy shot-to-shot. The average spectrum and $\pm1\sigma$ boundaries, overlaid as in (a), attains a maximum energy of 8.7~MeV, with peak single-shot fluxes consistently reaching $2\times10^{5}$~n/MeV/sr and selected shots reaching up to $5\times10^{5}$~n/MeV/sr. The higher cut-off energy of the neutron spectrum is enabled by the positive 4.36~MeV Q-value of the $^9$Be(d,n)$^{10}$B reaction. Note that the modulations near 1.5 and 2.2~MeV are artifacts in the raw nTOF detector signal, rather than features of the neutron spectrum. Their periodicity in the time domain suggests signal reflections in the cables between the nTOF detector and the oscilloscope. 

A simulated neutron spectrum was also obtained using the PHITS code as described above. The resulting spectrum is shown as a series of discrete points in \autoref{fig:neutron_energy_spectra}(b). While the simulated spectrum agrees with the experimentally measured spectrum within a factor of two for energies below 4~MeV, the simulation predicts elevated neutron fluxes relative to the experiment beginning around 5~MeV. This discrepancy may be due to a variety of factors, including reduced experimental nTOF resolution for high-energy neutrons due to the proximity of these signals to the decaying $\gamma$ flash signal in the raw nTOF trace. Additionally, the details of the simulated neutron spectra are determined by evaluated cross-section data, for which significant uncertainties remain in the description of deuteron-induced reaction channels \cite{Nakayama2021}. While detailed study of the exact reason for this discrepancy may be useful in benchmarking the code performance, further investigation is beyond the scope of this work.

\subsection{Directional neutron flux}
\label{sec:directionalflux}
The average forward-directed neutron flux can be calculated by averaging over the bubble detectors in the shadow of the converter or by integrating over the average spectrum detected by the two nTOF detectors. We note that we expect the modulations in the nTOF signal attributed to electronic noise to add only a small error to the total neutron flux calculation. As discussed below, the total neutron flux detected by the nTOF detectors agrees well with bubble detector measurements. We therefore infer that the contribution from the artifacts is within the shot-to-shot fluctuations of the nTOF signal. By calculating the average neutron flux obtained with and without the converter and with both ion species (protons or deuterons), we can compare the relative neutron generation performance between all four neutron generation conditions. Note that the measurement of bubble detectors for protons incident on the converter consisted of a measurement in which the converter was placed in the beam for 51\% of the recorded shots and removed for 49\% of the recorded shots, meaning the average flux value shown likely slightly underestimates the actual average neutron flux for protons incident on the converter. 

\begin{figure}
    \centering
    \includegraphics[width=\linewidth]{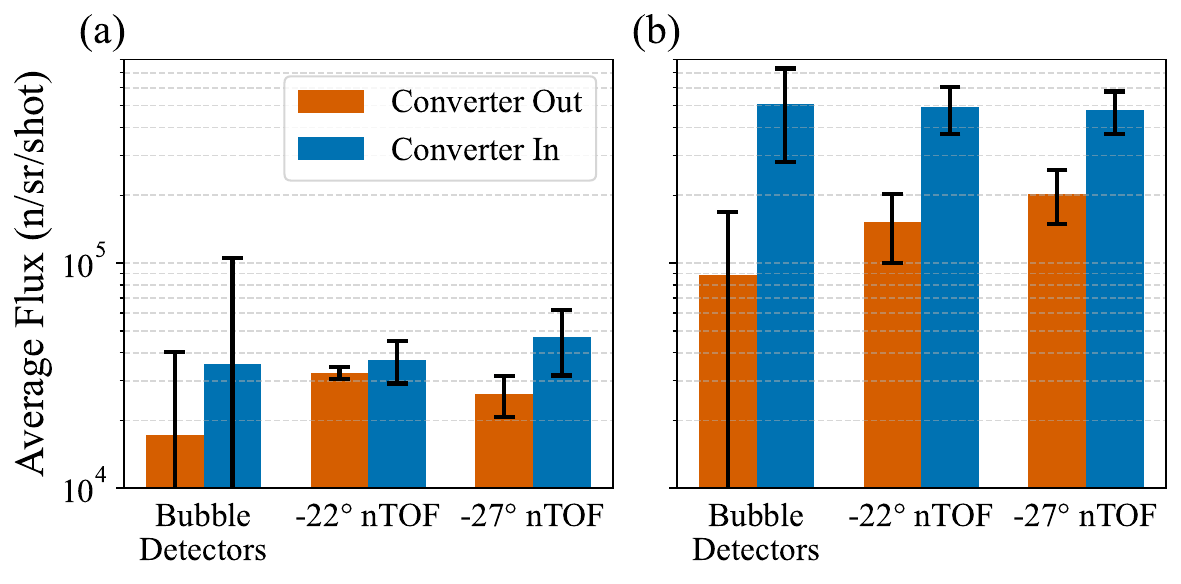}
    \caption{Measurements of the average forward-directed neutron flux in n/sr/shot obtained using both bubble detector dosimeters and the two nTOF detectors for incident protons (a) and deuterons (b). Both diagnostics record increased neutron fluxes when the converter is inserted in the ion beam, with a larger increase for deuterons than for protons. Error bars show the sample standard deviation. The highest average flux is obtained for a deuteron beam incident on the converter, reaching $\left(5.1\pm2.2\right)\times10^{5}$~n/sr/shot. Due to the nTOF calibration method, both nTOF detectors in (b) measure close to the same flux for the converter-in case.}
    \label{fig:p_d_neutrons}
\end{figure}

These results are presented in \autoref{fig:p_d_neutrons}. Both the bubble detectors and nTOFs observed the lowest neutron flux, calculated using the geometric center of the converter as the point source of neutrons, when a proton beam was used to generate neutrons without the converter. In this configuration, the neutron flux for the nTOF at -22$^\circ$ was below the noise threshold. When the converter was inserted into the proton beam, the single-shot directional neutron flux measured by the diagnostics increased by up to a factor of 2.1 as measured by the bubble detectors. This increase is due to neutron generation through the \textsuperscript{9}Be(p,n) reaction in the converter, which has a higher cross-section than the (p,n) reactions available with the Fe, Cr, Ni, and other trace elements composing the stainless steel chamber wall.

Neutron generation using a deuteron beam incident on the beryllium converter provided higher fluxes than using a proton beam, consistent with the addition of the deuteron breakup reaction to complete fusion reactions. When the converter was inserted into the deuteron beam, the neutron flux increased on average by a factor of 3.6 to the maximum value obtained in this experiment, $\left(5.1\pm2.2\right)\times10^{5}$~n/sr/shot as measured by the bubble detectors. We ascribe the larger increase for deuterons to the addition of the deuteron breakup channel, alongside the complete and incomplete fusion channels. The trends observed in the scaling of the neutron flux with the converter position and ion beam composition are consistent between both independently-calibrated nTOF and bubble detectors, providing further confidence in the observations.

\subsection{Experimental conversion efficiency}
In addition to comparing the forward-directed neutron flux produced by each ion species, we can also compare the corresponding estimates of total neutron yield. The experimental conversion efficiency was calculated by dividing the total neutron number by the estimated total ion number per shot. The total ion number generated can be approximated using the ion spectrum recorded by the Thomson parabola spectrometer while the converter was inserted and FWHM of the ion beam detected by the ion imager just prior to inserting the converter into the ion beam. For simplicity we assume that the ion beam divergence is independent of ion energy and we integrate the measured ion beam spectrum over the measured FWHM of the ion beam. The resulting total ion number per shot is shown in \autoref{tab:conversion_eff}. We note that, per \autoref{tab:ion_comparison}, the proton beam was offset from the TP axis by -3.3$^\circ$, while the deuteron beam was offset from the TP axis by -12.1$^\circ$. This means that the total proton number is likely a fairly accurate estimate, whereas the deuteron number is likely underestimated, because the TP sampled well off the deuteron beam axis. The total neutron number per shot was then approximated by integrating the average flux detected by all bubble detectors over 4$\pi$ assuming a uniform neutron emission into 4$\pi$. This approach averages out the forward-directed component of the neutron beam and therefore slightly underestimates the total generated neutron number, but is still useful in obtaining an order-of-magnitude conversion efficiency estimate for comparison between incident protons and deuterons.

\begin{table}
\caption{Total estimated ion number per shot based on Thomson parabola and ion imager data show a higher number for protons compared to deuterons. The opposite is true for the total number of neutrons generated by deuterons, which is more than an order of magnitude higher than the total number of proton-generated neutrons. The resulting conversion efficiency estimate from deuterons to neutrons is two orders of magnitude higher than that from protons to neutrons.}
\label{tab:conversion_eff}
\begin{ruledtabular}
\begin{tabular}{cccc}
\makecell{Ion\\species} & \makecell{Ions\\per shot} & \makecell{Neutrons\\per shot} & \makecell{Conversion\\efficiency} \\
\hline
Protons   & $1.7\times 10^{10}$ & $2.5\times 10^{5}$ & $1.5\times 10^{-5}$ \\
Deuterons & $5.3\times 10^{9}$  & $6.0\times 10^{6}$ & $1.1\times 10^{-3}$ \\
\end{tabular}
\end{ruledtabular}
\end{table}

By taking the ratio between neutron numbers per shot and ion numbers per shot we obtain a conversion efficiency from protons to neutrons of $1.5\times 10^{-5}$ and a conversion efficiency from deuterons to neutrons of $1.1\times 10^{-3}$, two orders of magnitude higher. Consistent with prior work, we conclude that the deuteron breakup channel significantly enhances the deuteron to neutron conversion efficiency compared to the proton to neutron conversion efficiency \cite{Roth2013}.

\section{Discussion}
This work presents a simultaneous spatial and spectral characterization of a HRR laser-driven ion beam and laser-driven neutron source in the pitcher-catcher geometry. After obtaining a detailed characterization of the laser-accelerated deuteron beam driving the neutron source, measurements of the laser-generated neutrons using both bubble detectors and nTOF detectors recorded an average flux across the different diagnostics of approximately $4.9\times10^{5}$~n/sr/shot with a deuteron to neutron conversion efficiency of $1.1\times 10^{-3}$ calculated from averaging over more than 200 consecutive shots. Measurements of the forward-directed neutron flux at different distances from the converter showed that neutrons were predominantly produced in the converter when it was placed in the ion beam. The spectrum of the neutrons detected at an angle of -27$^\circ$ exhibited a broadband energy spectrum with an average maximum energy of 8.7~MeV. While we found good agreement between nTOF measurements and PHITS simulations for peak and average neutron fluxes, clear discrepancies were visible in the spectral shape, especially for energies above 5~MeV. This may be the result of the simplified model of the experimental setup used in the simulation, as other hardware in the densely populated chamber could have provided a neutron generation or scattering source. 

When investigating the neutron yield driven by proton versus deuteron beams, we observed an increase in average flux of around one order of magnitude when switching from protons ($4\times 10^{4}$ n/sr/shot) to deuterons  ($5.1\times 10^{5}$ n/sr/shot). This corresponds to an increase in conversion efficiency from ions to neutrons of two orders of magnitude, which we attributed to the additional neutron generation pathway of deuteron breakup. Peak integrated neutron fluxes achieved when deuterons were incident on the converter reached $6\times 10^{6}$~n/shot, corresponding to $3\times 10^{6}$~n/s at a repetition rate of 0.5~Hz and $1.25\times 10^{6}$~n/shot/J of incident laser energy. These observations are consistent with previous measurements of HRR pitcher-catcher laser-driven neutron sources, but feature the key advance of simultaneous on-shot measurements of both the deuteron and neutron energy spectra. 

Previous work demonstrating similar fluxes at the 0.5~Hz level relied on bubble detectors and activation detectors to obtain neutron fluxes integrated across the entire shot series \cite{Lelievre2024}. In another experiment demonstrating neutron generation from a kHz laser system delivering pulse energies of tens of mJ, single-shot neutron fluxes were too low to be measured on the nTOF detectors and the data presented are integrated over thousands of shots \cite{Stuhl2025}. While the neutron yield in this study is lower than recently published results on electron-driven neutron generation by Valli\`{e}res et al. \cite{Vallieres2025}, we present several avenues to improve the overall neutron yield of our platform in the conclusion. The direct correlation between ion beam and neutron beam parameters also gives ion-driven neutron sources a larger parameter space to adapt neutron beam parameters to the desired application. The measurements here demonstrate this capability and permit the shot-by-shot neutron spectra to be associated to the driving deuteron beam parameters, facilitating detailed comparison to simulations. Future uses of laser-driven neutron sources in applications such as neutron resonance spectroscopy will also benefit from continuous monitoring of the on-shot neutron energy spectra, allowing sensitive measurements to directly incorporate measured variations in the neutron source spectrum as well as the implementation of active feedback loops based on the collected data to optimize or stabilize the ion and neutron beam output parameters.

\section{Conclusion}
We developed an ion-driven neutron generation platform with simultaneous ion and neutron beam monitoring for advanced control and optimization concepts towards applications. The neutron fluxes presented here are already suitable for static radiography measurements in which flux can be accumulated over many shots. This results from not only the HRR nature of the source, which allows hundreds of shots to be accumulated in just a few minutes, but also the demonstrated stability of the source on these timescales. The liquid sheet target is inherently compatible with repetition rates up to the kHz level \cite{Treffert2022a}, which will facilitate significant increases in the achievable average neutron flux in future work. Demanding future applications such as fast neutron spectroscopy or single-shot radiography of driven samples will require significant increases in the peak neutron flux and peak energy. Beyond simply increasing the laser energy and power (which has previously been shown to yield significantly higher neutron yields than those shown here \cite{Kleinschmidt2018}), improvements in the peak neutron flux and peak energy can be achieved by rotating the laser polarization to p-polarization for improved laser-target coupling and modifying the laser wavelength from 400~nm to 800~nm  \cite{Keppler2022,Wilks2025}. Alternative liquid nozzle types, such as converging glass nozzles \cite{Crissman2022}, may also enable the generation of thinner liquid sheets to further improve the ion yield and maximum energy. Recent observations of remarkably low-divergence ion beams generated by liquid sheet targets \cite{Streeter2025} may also provide an opportunity to obtain laser-driven neutron sources with a significantly reduced source size and increased flux. Finally, recent demonstrations of closed-loop machine learning optimization \cite{Loughran2023,Glenn2026} of laser-driven ion beams offer the promise of both active stabilization of laser-driven neutron sources and online tuning of the neutron source flux or spectrum as required for applications. This collection of techniques provides a clear roadmap from the platform demonstrated here toward experiments which will demonstrate application-ready laser-driven neutron sources delivering high single-shot fluxes with in-situ adjustable spectra at HRR.

\begin{center}\textbf{Acknowledgements}\end{center}
\noindent F.T., G.D.G, C.B.C., G.J., M.R., M.G., and S.H.G. acknowledge support from the U.S. DOE Office of Science, Fusion Energy Sciences under FWP Nos.~100182 and 100866 and in part by the NSF under Grant Nos.~1632708 and PHY-2308860. Work was performed at LLNL under Contract No. DE-AC52-07NA27344 and supported by the LDRD program under tracking Code No. 22-ERD-022. G.D.G. acknowledges support from the DOE NNSA SSGF program under DE-NA0003960. Sandia National Laboratories is a multi-mission laboratory managed and operated by National Technology \& Engineering Solutions of Sandia, LLC (NTESS), a wholly owned subsidiary of Honeywell International Inc., for the U.S. Department of Energy’s National Nuclear Security Administration (DOE/NNSA) under contract DE-NA0003525. This paper describes objective technical results and analysis. Any subjective views or opinions that might be expressed in the paper do not necessarily represent the views of the US Department of Energy or the United States Government. Use of the Linac Coherent Light Source (LCLS), SLAC National Accelerator Laboratory, is supported by the U.S. Department of Energy (DOE), Office of Science, Office of Basic Energy Sciences (BES) under Contract No. DE-AC02-76SF00515. This work was supported by the U.S. Department of Energy’s (DOE) Office of Science (SC) Fusion Energy Sciences (FES) program under DE-SC0021246: the LaserNetUS initiative at the Advanced Beam Laboratory. R.H. acknowledges support from the U.S. DOE Office of Science, Fusion Energy Sciences under Award Number DE-SC0025450. S.P. and D.U. acknowledge support from the project ``Medical applications of high-power lasers -- Dr. LASER'' -- SMIS Code: 326475 funded by the European Union, the Romanian Government and the Health Program and project ELI-RO/DFG/2025-013 IATP-NP 2.0 funded by the Institute of Atomic Physics, Romania. J.J.R. acknowledges support from the DoD Vannevar Bush Faculty Fellowship, ONR Award N000142012842.

\begin{center}\textbf{Data Availability Statement}\end{center}
The data that support the findings of this study are available from the corresponding author upon reasonable request.

\bibliographystyle{apsrev4-2}
\bibliography{Treffert_Glenn_2026_References_arXiv}

\end{document}